\documentclass[aps,prl,twocolumn,showpacs,superscriptaddress,groupedaddress]{revtex4}
\usepackage{mathbbol}
\usepackage{amsmath}
\usepackage{amsfonts}
\usepackage{amssymb}
\usepackage{graphicx}
\usepackage{morefloats}
 \usepackage{dcolumn}   
\usepackage{bm}       
\usepackage{amssymb}

\begin{document}

\newcommand{\ket}[1]{\left |#1 \right \rangle}
\newcommand{\bra}[1]{\left  \langle #1 \right |}
\newcommand{\braket}[2]{\left \langle #1 \middle | #2 \right \rangle}
\newcommand{\ketbra}[2]{\left | #1 \middle \rangle \middle \langle #2 \right |}

\widetext

\title{Symmetric 3 Qubit State Invariants}
\author{Alexander Meill and David A. Meyer}
\date{\today}

\begin{abstract}
For pure symmetric 3-qubit states there are only three algebraically independent entanglement measures; one choice is the pairwise concurrence $\mathcal C$, the 3-tangle $\tau$, and the Kempe invariant $\kappa$.  Using a canonical form for symmetric $n$ qubit states derived from their Majorana representation, we derive the explicit achievable region of triples $(\mathcal C,\tau,\kappa)$.
\end{abstract}

\maketitle

\section{Introduction}
Entanglement is a critical resource for quantum computational tasks such as teleportation \cite{1} and cryptography \cite{2} among many others, but analytic calculations of entanglement quickly become challenging as the dimension of the Hilbert space increases.  Restricting those calculations to states which are symmetric under subgroups of permutations of party labels can greatly reduce the number of parameters and simplify the calculations considerably.  Symmetric states are particularly useful for this reason; in addition, symmetric states are relevant in many experimental settings such as in Measurement-Based Quantum Computing \cite{3}, as an initial state for Grover's Algorithm \cite{4}, and as ground states of various translation invariant Hamiltonians \cite{17}.  In this paper, the restriction to completely symmetric states is used to simplify the calculation of 3 qubit local unitary (LU) invariants.

The entanglement of an $n$ qubit state, as defined by any measure, remains invariant under local unitary operators of the form $U=U_1 \otimes U_2 \otimes \ldots \otimes U_n$, where $U_i \in U(2)$.  Polynomial invariants of a multi-qubit state are not limited to entanglement monotones, though these are a particularly relevant choice.  In this paper, the invariants of an arbitrary three qubit symmetric state are calculated explicitly and their full achievable region is calculated.

\section{3 Qubit Polynomial Invariants}
Any multi-particle state has a set of polynomials in the coefficients of the state which are invariant under the action of various local operators \cite{5}.  In particular, a 3 qubit state, under the action of unitaries which act only locally on one qubit, is known to have 5 algebraically independent invariants (as well as the trace norm and $\mathbb{Z}_2$ invariant) \cite{6}.  There is some freedom in choosing 5 generators of the algebra of invariant polynomials, as any polynomial in invariants is additionally an invariant of the state.  One set of generators that is particularly convenient for 3 qubit states under local unitary operators is
\begin{equation}
\left \{ \mathcal{C}_{1,2}, \, \mathcal{C}_{2,3}, \, \mathcal{C}_{3,1}, \, \tau, \, \kappa \right \},
\end{equation}
where $\mathcal{C}_{i,j}$ is the pairwise concurrence between parties $i$ and $j$ \cite{7}, $\tau$ is the 3-tangle \cite{8}, and $\kappa$ is the Kempe invariant \cite{9}.  These quantities are defined for a 3 qubit state, $\ket{\psi} = \sum_{i,j,k=0}^1 c_{ijk} \ket{ijk}$, as follows:
\begin{equation}
\mathcal{C}_{i,j} = \max\left\{0,\lambda_1-\lambda_2-\lambda_3-\lambda_4\right\},
\end{equation}
\begin{equation}
\begin{split}
\tau = 2 \LARGE{|} & \epsilon_{i_1 i_2} \epsilon_{i_3 i_4} \epsilon_{j_1 j_2} \epsilon_{j_3 j_4}  \epsilon_{k_1 k_3}  \epsilon_{k_2 k_4}  \\ & \times c_{i_1 j_1 k_1} c_{i_2 j_2 k_2} c_{i_3 j_3 k_3} c_{i_4 j_4 k_4} \LARGE{|},
\end{split}
\end{equation}
\begin{equation}
\kappa = c^{\hphantom *}_{i_1 j_1 k_1}  c^{\hphantom *}_{i_2 j_2 k_2} c^{\hphantom *}_{i_3 j_3 k_3} c_{i_1 j_2 k_3}^*  c_{i_2 j_3 k_1}^* c_{i_3 j_1 k_2}^* ,
\end{equation}
where the $\lambda_\alpha$ in (2) are the square roots of the eigenvalues, in decreasing order, of $\rho_{i,j} (\sigma_y \otimes \sigma_y) \rho_{i,j}^* (\sigma_y \otimes \sigma_y)$.  Here, $\rho_{i,j}$ is the reduced density operator of the state having traced out all parties other than $i$ and $j$.  Note, also, that in (3) and (4) we have adopted the convention of summing over repeated indices.  This choice of invariants is particularly useful as it uses some of the most prevalent entanglement measures in the concurrence and 3-tangle.

It would be interesting to completely map the space of these 5 invariants, but the calculations are difficult for arbitrary states, and the 5 dimensional picture would be unwieldy to describe or visualize.  Instead, by examining a particular subset of states, the number of parameters and invariants can be reduced to be more manageable.  In the next section this is done for the subset of states which are symmetric under permutation of the party labels.

\section{Symmetric 3 Qubit States}

Symmetric states offer a significant simplification to the picture of 3 qubit invariants.  Clearly if a state is symmetric under relabeling of parties, each of the two-party reduced density operators, $\rho_{i,j}$, will be identical.  This then causes $\mathcal{C}_{1,2}=\mathcal{C}_{2,3}=\mathcal{C}_{3,1}=\mathcal{C}$ and effectively reduces the number of invariants to 3, which will be denoted,
\begin{equation}
\left \{ \mathcal{C}, \tau, \kappa \right \}.
\end{equation}
The goal now is to calculate these invariants for a general 3 qubit symmetric state and describe the region of allowed and acheivable values for these invariants.  Before doing so, however, we can further simplify the problem by examining how the symmetric subspace of 3 qubit states reduces the parameters on which the invariants depend.

The most natural representation of an $n$ qubit symmetric state is in the Dicke basis \cite{10},
\begin{equation}
\ket{\psi} = \sum_{i=0}^n a_i \ket{S_i^{(n)}},
\end{equation}
where
\begin{equation}
\ket{S_i^{(n)}}={n \choose i}^{-1/2} \sum_{\pi \in S_n} \pi | \underbrace{00...0}_{n-i}\underbrace{11...1}_{i} \rangle
\end{equation}
are the Dicke basis states which represent an equal superposition of all possible states with $i$ ``0'' entries and $n-i$ ``1'' entries.  They are obviously symmetric since $\pi$ permutes the parties of the state and the sum is an equal superposition of all possible permutations.  In this representation, the 3 qubit symmetric state has 4 complex coefficients, $a_i$, which reduce to 6 real parameters after normalization and the factoring out of an overall phase.  While the invariants can be calculated from these 6 real parameters, it is useful to apply a set of local unitaries to the state to reduce the number of parameters.  The calculation of the invariants then becomes more concise while still containing the same information since the invariants should not change under local unitaries.  Such a simplification was investigated in [11].  Using almost the same argument, we show in the following that most 3-qubit symmetric states are equivalent under local unitaries to states of the form,
\begin{equation}
\ket{\psi'} = A \left(\ket{000} + y e^{i \phi} \ket{\theta} ^{\otimes 3} \right),
\end{equation}
where $y\in[0,1)$, $\theta\in[0,\pi]$, $\phi\in[0,2\pi)$,
$|\theta\rangle = \cos(\theta/2)|0\rangle + \sin(\theta/2)|1\rangle$ is a single qubit
state with purely real coefficients, and $A$ is a normalization constant.  What follows is a proof of (8).

Consider an arbitrary 3 qubit symmetric state given by,
\begin{equation}
\ket{\psi} = \sum_{i=0}^3 a_i \ket{S_i^{(3)}}.
\end{equation}
One can compute the Majorana Polynomial \cite{12} of $\ket{\psi}$, by projecting it onto the unnormalized state $\ket{\alpha} = \left( \ket{0} + \alpha^* \ket{1} \right)^{\otimes 3}$, where $\alpha$ is an arbitrary complex number.  The resulting inner product is the following polynomial in $\alpha$,
\begin{equation}
\braket{\alpha}{\psi} = a_0 + \sqrt{3} a_1 \alpha + \sqrt{3} a_2 \alpha^2+ a_3 \alpha^3.
\end{equation}

By the first fundamental theorem of algebra, the roots of this polynomial, $\alpha_i$, known as the Majorana roots, uniquely determine the coefficients $a_i$ up to some overall scaling factor.  So to show that two states are the same, it suffices to show that they have the same Majorana roots.  We will exploit this fact to show that most states, $\ket{\psi}$, can be represented as
\begin{equation}
\ket{\psi} = c_0 \ket{\Phi_0} + c_1 \ket{\Phi_1},
\end{equation}
where $\ket{\Phi_j}=\left( \cos \theta_j \ket{0} + e^{i \phi_j} \sin \theta_j \ket{1} \right)^{\otimes 3}$ and $c_j \in \mathbb{C}$, by showing that (11) has the same Majorana roots as (9) for some choice of $c_j$, $\theta_j$ and $\phi_j$.  Before computing the Majorana roots of (11), we can factor out $c_0$ from the state to express it as,
\begin{equation}
\ket{\psi} = A\left( \ket{\Phi_0} + c \ket{\Phi_1} \right),
\end{equation}
where $A$ is a normalization constant and $c = c_1/c_0 \in \mathbb{C}$.  The Majorana polynomial of (12) can be expressed as
\begin{equation}
\braket{\alpha}{\psi} = \left( \cos \theta_0 + \alpha \sin \theta_0 e^{i \phi_0} \right)^3 + c \left( \cos \theta_1 + \alpha \sin \theta_1 e^{i \phi_1} \right)^3.
\end{equation}
Note we have dropped the normalization factor, $A$, since we need only specify the polynomial up to a scaling factor.  We can further simplify by dividing by $\cos \theta_0$, which leaves,
\begin{equation}
\braket{\alpha}{\psi} = \left( 1 + \alpha \beta_0 \right)^3 + c' \left( 1 + \alpha \beta_1 \right)^3,
\end{equation}
where $\beta_j = \tan \theta_j \, e^{i \phi_j}$ and $c' = c (\cos \theta_1)/(\cos \theta_0)$. The goal is to demonstrate that one can solve for a choice of $\beta_j$ and $c'$ that will satisfy (14) having the same roots as (10).  This creates the following constraints on $\beta_j$ and $c'$,
\begin{eqnarray}
0 &=& \left( 1 + \alpha_1 \beta_0 \right)^3 + c' \left( 1 + \alpha_1 \beta_1 \right)^3 \\
0 &=& \left( 1 + \alpha_2 \beta_0 \right)^3 + c' \left( 1 + \alpha_2 \beta_1 \right)^3.
\end{eqnarray}

Additionally, we can require that the projection of (12) onto $\ket{\alpha}$ be the same as (10) when evaluated at $\alpha=0$, which provides the third constraint,
\begin{equation}
a_0 = c_0 \left(\cos \theta_0 \right)^3 + c_1 \left(\cos \theta_1 \right)^3.
\end{equation}

Equations (15-17) provide sufficient constraints on $\beta_j$ and $c'$ to identify a representation (11) which is the same state as (9) so long as no Majorana root, $\alpha_i$ is degenerate with degree 2 \cite{11}.  One can then construct a local unitary operator of the form $\mathcal{U} = U \otimes U \otimes U$, where $U \left( \cos \theta_0 \ket{0} + \sin \theta_0 e^{i \phi_0} \ket{1} \right) = \ket{0}$.  Applying this to (12) results in
\begin{equation}
\mathcal{U} \ket{\psi} = A \left( \ket{000} + c \ket{\chi} \right),
\end{equation}
where $\ket{\chi} = \mathcal{U} \ket{\Phi_1} = \left( \cos \theta/2 \ket{0} + \sin \theta/2 e^{i \chi} \ket{1} \right)^{\otimes 3}$.  Lastly, one can apply the local unitary operator,
\begin{equation}
\mathcal{V} = \left(\begin{array}{cc}1 & 0 \\0 & e^{-i \chi}\end{array}\right) ^ {\otimes 3}
\end{equation}
to the state (18) to arrive finally at the desired result,
\begin{equation}
\ket{\psi'} = A \left(\ket{000} + y e^{i \phi} \ket{\theta} ^{\otimes 3} \right),
\end{equation}
where $\ket{\theta} = \cos \theta/2 \ket{0} + \sin \theta/2 \ket{1}$, and $c = y e^{i \phi}$.  

This canonical form is a useful tool in examining the local unitary invariants of 3-qubit symmetric states.  Other canonical forms and representations for 3-qubit states are known, but this one provides considerable simplification to symmetric states in particular and leaves them symmetric after the rotation.  If one were to transform (20) into one of the general 3-qubit canonical forms developed by Ac{\'\i}n et al. in \cite{13} one would get the following state,
\begin{equation}
\begin{split}
\ket{\chi_1} = A\biggr{[} \sin \theta \ket{000} + \cos \theta \left(e^{-i \theta}+ y \cos \theta \right) \ket{100} \\
 + y \cos \theta \sin \theta \left( \ket{101} + \ket{110} \right) + y \sin^2\theta \ket{111} \biggr{]},
\end{split}
\end{equation}
which is less obviously symmetric than (20).  An effective canonical form utilizing the inner products of the vectors in the Majorana representation was presented in \cite{14}. This form was used to calculate a slightly different set of invariants, $ \left \{ \tau, \kappa, \text{Tr}(\rho_1^2) \right \}$, where $\rho_1$ is the single-party reduced density matrix of the symmetric state obtained by tracing out any two parties, and the states which maximize and minimize the various invariants were examined.  In what follows, we will calculate the $\left \{ \mathcal C, \tau, \kappa \right \}$ of (20) and describe the full achievable space of those variants.

In terms of the parameters $y$, $\theta$, and $\phi$ of (20), the invariants are,
\begin{eqnarray}
\tau &=& \frac{2y\sin^3\frac{\theta}{2}}{1+y^2+2y\cos^3\frac{\theta}{2}\cos\phi} \\
\mathcal{C} &=& \frac{y\sin\frac{\theta}{2}\sin\theta}{1+y^2+2y\cos^3\frac{\theta}{2}\cos\phi}
\end{eqnarray}
\begin{equation}
\begin{split}
\kappa =& \frac{1}{8\left(1+y^2+2y\cos^3\frac{\theta}{2}\cos\phi \right)} \times \\
\biggr{[}& \left(1+y^2 \right)
\left( 8+19y^2 + 8 y^4 + 9y^2 \left(4 \cos \theta + \cos 2\theta \right) \right)  \\
+&24y\cos^3\frac{\theta}{2} \left( 2+3y^2+2y^4+3y^2\cos \theta \right) \cos\phi \\
 + &48y^2 \left( 1+y^2\right) \cos^6 \frac{\theta}{2}\cos2\phi+16y^3\cos^9\frac{\theta}{2} \cos3\phi \biggr{]}.
\end{split}
\end{equation}

Figure 1 shows the invariants of $10^5$ randomly generated symmetric 3 qubit states, where the states were generated by sampling randomly over the allowed values of $y$, $\theta$, and $\phi$.
\begin{figure}[h]
\centering
\includegraphics[width = 80mm]{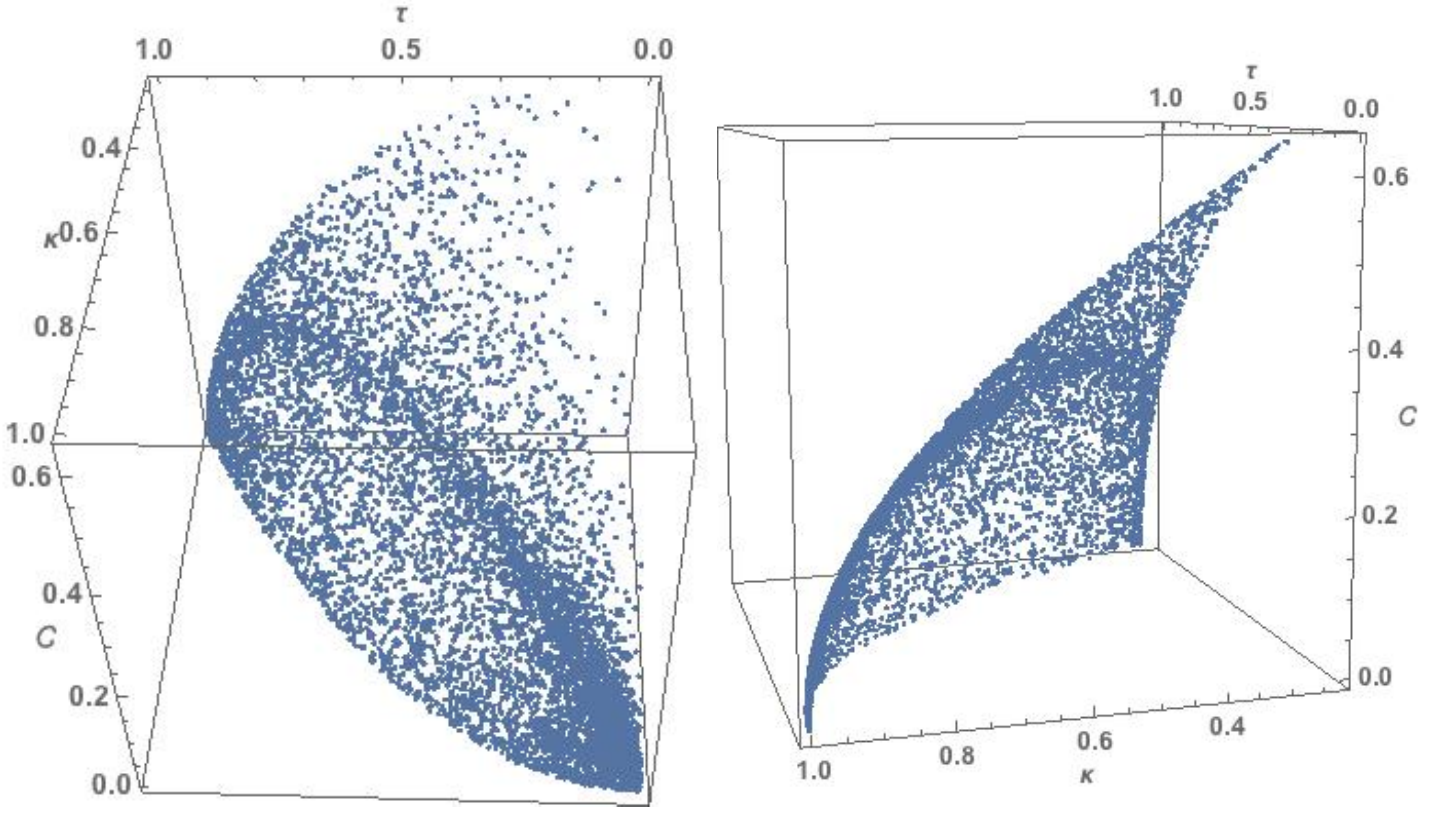}
\caption{Scatterplot from two points of view of invariants of randomly generated symmetric 3 qubit states.}
\end{figure}
At a first glance, it is interesting to note that the 3-tangle and Kempe invariants achieve their maximum values of 1 on the symmetric subspace, but the concurrence does not due its monogamy constraints \cite{8}.  A straightforward maximization over the state parameters reveals a maximum concurrence of $2/3$ in the symmetric subspace, which confirms the result of \cite{15} for $n=3$.  The points of Figure 1 appear to lie almost on a surface, but closer inspection reveals that they in fact fill a narrow volume, the boundaries of which can be calculated.  We can invert the expressions (22-24) by a Gr\"obner basis calculation to find,
\begin{eqnarray}
\cos \frac \theta 2 &=& \frac{\mathcal C}{\sqrt{\mathcal C^2 + \tau^2}} \\ 
\cos \phi &=& \frac{ 4-3\tau^2-9\mathcal C^2-4\kappa}{3\mathcal C^3}
\end{eqnarray}
\begin{equation}
\begin{split}
y =&\frac{6\tau^2+9\mathcal C^2+4\kappa -4}{3 \left(\tau^2+\mathcal C^2\right)^{3/2}} \\
&-\sqrt{\left( \frac{6\tau^2+9\mathcal C^2+4\kappa -4}{3 \left(\tau^2+\mathcal C^2 \right)^{3/2}} \right)^2-1}.
\end{split}
\end{equation}
The constraints on the state parameters then provide constraints on these functions of the invariants.  The extrema of these constraints are the surfaces which form the boundaries of the invariant space.  The boundaries are formed when equality is achieved in the following relations.
\begin{eqnarray}
0 &\leq& 4-\tau^2-9\mathcal{C}^2-4\kappa+3\mathcal{C}^3 \\
0 &\geq& 4-\tau^2-9\mathcal{C}^2-4\kappa-3\mathcal{C}^3 \\
0 &\geq& 4-6\tau^2-9\mathcal{C}^2-4\kappa +3\left(\tau^2+\mathcal{C}^2\right)^{3/2}. 
\end{eqnarray}

\begin{figure}[h]
\centering
\includegraphics[width = 60mm]{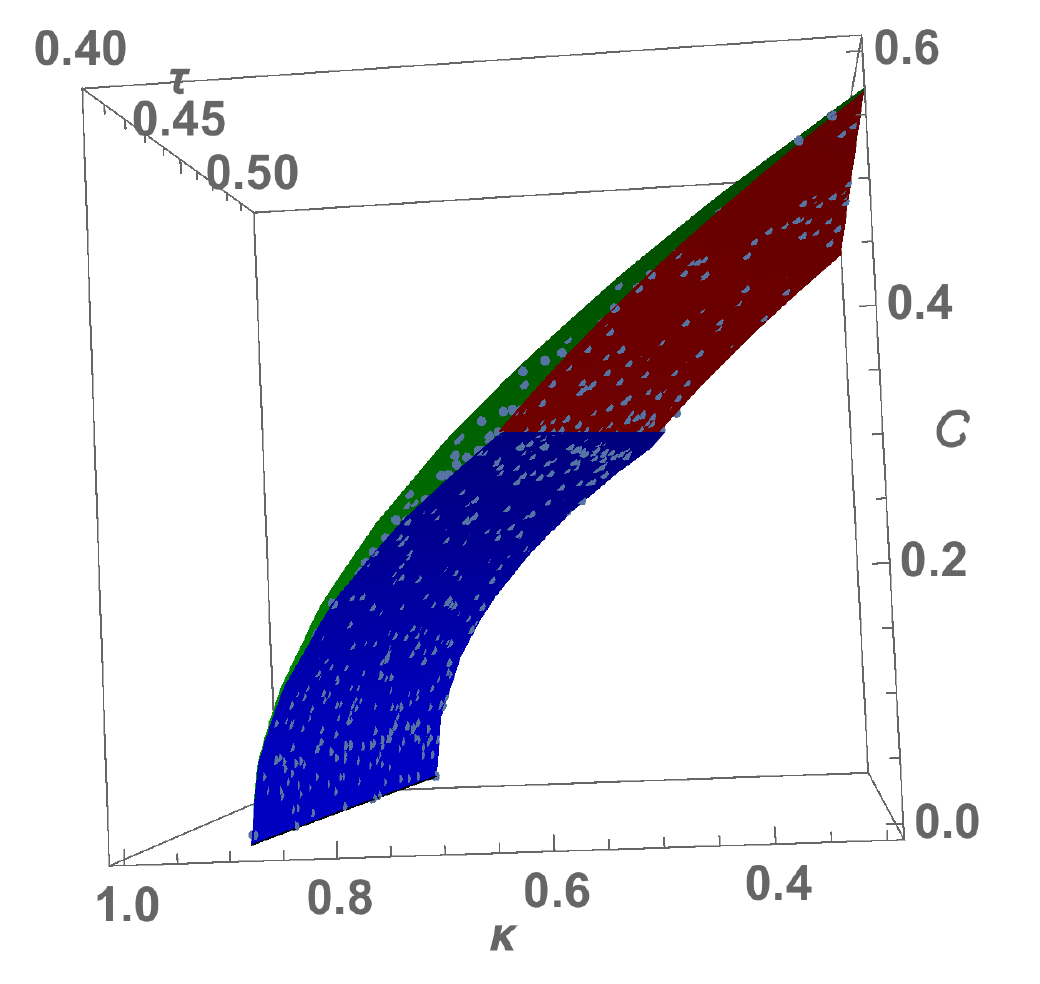}
\caption{View of a slice of the boundaries of the volume of symmetric 3 qubit invariants superimposed over the points of Figure 1.  The contour achieving equality in equation (28) is shown in green, (29) in blue, and (30) in red.}
\end{figure}

These three surfaces, which are shown in Figure 2, form boundaries for the possible space of the invariants and serve as additional monogamy relations for symmetric 3 qubit entanglement.  Note that the state parameter constraints lead to more constraints on the invariants, but (28-30) is the minimum set of constraints required to describe the region.  Because there is a bijective map between the invariants and the state parameters, each invariant triple which lies within the region satisfying (28-30) can be mapped to a 3 qubit symmetric state, and therefore the entire region is achievable.

We should at this point address the states which do not admit a representation of the form (8), which we denote $\ket{\bar{\psi}}$.  It is shown in \cite{11} that 3 qubit states which have a degenerate Majorana root with degree 2 cannot be expressed in this canonical form.  Instead, we will parametrize states of that form and show that the invariants of this subset of state likewise satisfy (28-30).  An arbitrary 3 qubit state with a degenerate Majorana root of degree 2 can be expressed in the Majorana representation as
\begin{equation}
\ket{\bar{\psi}} = \frac 1A\sum_{\pi \in S_3} \pi \ket{\phi_1} \otimes \ket{\phi_1} \otimes \ket{\phi_2},
\end{equation}
where $\phi_i$ are single qubit states.  We can again use local unitaries to simplify states of this form to
\begin{equation}
\ket{\bar{\psi}'} = \frac 1A\sum_{\pi \in S_3} \pi \ket{0} \otimes \ket{0} \otimes \ket{\theta},
\end{equation}
where $\ket{\theta}$ is the same as in (8) for $\theta \in (0,\pi]$.  The invariants of (32) are
\begin{eqnarray}
\tau &=& 0, \\
\kappa &=& \frac{2+48 \cos^2 \frac {\theta}2 +141 \cos^4 \frac {\theta}2 + 52 \cos^6 \frac {\theta}2}{9(1+2 \cos^2 \frac {\theta}2)^3}, \\
\mathcal C &=& \frac{2-2 \cos^2 \frac {\theta}2}{3+6 \cos^2 \frac {\theta}2}.
\end{eqnarray}
It is then easy to verify that (33-35) satisfy (28-30) for $\theta \in (0,\pi]$.  This is perhaps unsurprising given that the states with a degenerate Majorana root of degree 2 are a limiting case of states which admit the canonical form.  Now all 3 qubit symmetric states have been considered and it can be concluded that (28-30) do indeed describe the full achievable region for 3 qubit symmetric states.

A similar approach could be used to analyze the invariants of $n$ qubits for the symmetric subspace.  The representation in \cite{11} extends to symmetric $n$ qubit states resulting in $2n-3$ real parameters.  3 qubits, in particular, can be fully analyzed and visualized because the number of invariants and state parameters is suitably low.  Additionally, in the 3 qubit case, remarkably there is an invertible map between the state parameters and the invariants, allowing our the calculation of the achievable region.  Turning to the $n>3$ qubit case, \cite{14} and \cite{16} use the Majorana state representation to examine the SLOCC classes and invariants of symmetric states, but the LU invariants remain less explored.  The inner products of the vectors in the Majorana representation are themselves a set of $2n-3$ LU invariants, as used by \cite{14}.  It would be interesting to find an alternate set of $2n-3$ algebraically independent LU invariants which includes pertinent entanglement measures.  That set of invariants could potentially then be calculated in terms of the $2n-3$ state parameters.  The remarkable fact that this map was invertible for 3 qubits will not necessarily be true for the $n$ qubit case, though it is certainly worth examining.

\section*{Acknowledgements}

We thank the authors of \cite{14} for bringing their work to our attention, and NSF grant PHY-1620846 for partial support of this research.

\end{document}